\documentstyle[12pt,epsfig]{article}

\def \sect #1 {\setcounter{equation} 0\section{#1}}
\def \be  {\begin{equation}}
\def \ee  {\end{equation}}
\def \ba  {\begin{eqnarray}}
\def \ea  {\end{eqnarray}}
\def \baa {\begin{eqnarray*}}
\def \eaa {\end{eqnarray*}}
\def \bb  {\begin{thebibliography}}
\def \eb  {\end{thebibliography}}

\def \lab #1 {\label{#1}}

\def \CO {{\cal O}}

\def \fracs #1#2 {\mbox{\small $\frac{#1}{#2}$}}

\def \bin #1#2 {{\left({#1}\atop{#2}\right)}}
\def\lapproxeq{{\ \lower 0.6ex \hbox{$\buildrel<\over\sim$}\ }}
\def\gapproxeq{{\ \lower 0.6ex \hbox{$\buildrel>\over\sim$}\ }}

\def \as {\relax\ifmmode\alpha_s\else{$\alpha_s${ }}\fi}

\def \al #1 {\frac {\as({#1})}{\pi} }
\def \ds #1 {\ooalign{$\hfil/\hfil$\crcr$#1$}}

\def \prt{perturbative }

\def \CO {{\cal O}}
\def\np #1#2#3  {{Nucl. Phys.~{\bf #1} (19#3) #2 }}
\def\nc #1#2#3  {{Nuovo. Cim.~{\bf #1} (19#3) #2}}
\def\pl #1#2#3  {{Phys. Lett.~{\bf #1} (19#3) #2}}
\def\pr #1#2#3  {{Phys. Rev.~{\bf #1} (19#3) #2}}
\def\prl #1#2#3  {{Phys. Rev. Lett.~{\bf #1} (19#3) #2}}
\def\prep #1#2#3 {{Phys. Rep.~{\bf #1} (19#3) #2}}
\def\zp #1#2#3  {{Z. Phys.~{\bf #1} (19#3) #2}}
\def\epj #1#2#3  {{Eur. Phys. J.~{\bf #1} (#3) #2}}
\def\rmp #1#2#3  {{Rev. Mod. Phys.~{\bf #1} (19#3) #2}}
\def\JETP #1#2#3 {{Sov.\ Phys.\ JETP~{\bf #1} (19#3) #2}}
\def\sj #1#2#3 {{Sov.\ J.\ Nucl.\ Phys.~{\bf #1} (19#3) #2}}
\def\hepph  #1 {{hep-ph/#1 }}

\begin{document}

\begin{flushright}
BNL-HET-03/20 \\
BNL-NT-03/26 \\
RBRC-335  \\
YITP-SB-03-47 \\
\today
\end{flushright}

\vspace*{10mm}

\begin{center}
{\LARGE \bf
Joint Resummation for Higgs Production}

\par\vspace*{10mm}\par

{\large Anna Kulesza$^a$,
George Sterman$^b$, Werner Vogelsang$^c$}

\bigskip

{\em $^a$Department of Physics, Brookhaven National Laboratory,
Upton, NY 11973, U.S.A.}

\bigskip

{\em $^b$C.N.\ Yang Institute for Theoretical Physics,
Stony Brook University \\
Stony Brook, New York 11794 -- 3840, U.S.A.}

\bigskip

{\em $^c$RIKEN-BNL Research Center and Nuclear Theory, \\
Brookhaven National Laboratory,
Upton, NY 11973, U.S.A.}

\end{center}
\vspace*{15mm}

\begin{abstract}
\noindent
We study  the application of the joint resummation formalism to Higgs
production via gluon-gluon fusion at the LHC, defining inverse  transforms
by analytic continuation.   We work at next-to-leading logarithmic 
accuracy. We find that at low $Q_T$ the resummed Higgs $Q_T$ distributions
are comparable in the joint and pure-$Q_T$ formalisms,
with relatively small influence from threshold enhancement
in this range.   We find a modest (about ten percent)
decrease in the inclusive cross section, relative to pure threshold 
resummation.
\end{abstract}

\newpage
\section{Introduction}

The gluon-gluon fusion process $gg \rightarrow hX$ is expected to be
the dominant mechanism for the production of rather light Higgs bosons 
at
the LHC~\cite{LesHouches02}.
Next-to-leading order (NLO) QCD corrections to the total production
cross section
for this process are known to be large, of the order of 70\%~\cite{NLO}.
This made the computation of NNLO corrections to the process an
important task, which was completed very recently~\cite{NNLO}.
The NNLO  corrections are found to be substantial as well,
even at LHC energies, where they increase the total cross section 
by an additional 30\%~\cite{NNLO} over NLO.
The dominant part of these contributions is directly related to soft and
collinear gluon emission, as demonstrated
in Refs.~\cite{NNLOsoftcolCdFG,NNLOsoftcolHK}.

Soft gluon emission is manifested through the presence of
logarithmic corrections in expressions for partonic subprocesses.
These corrections arise from cancellations between real and virtual
contributions to the differential cross section at each order in
perturbation theory. They can be large when a
cross section is sensitive to their behavior near the boundary of phase 
space. This is the case for threshold corrections where terms of the 
form \mbox{$\as^n \ln^{2n-1}(1-z)/(1-z)$} grow large in the limit 
$z=Q^2/\hat s \rightarrow 1$, that is, when the partonic center-of-mass
energy $\hat s$ approaches the invariant mass $Q$
of the Higgs particle. Similarly, recoil corrections appearing in
transverse momentum distributions, $\as^n \ln^{2n-1}(Q^2/Q_T^2)/Q_T^2$, 
become large when the Higgs boson is produced
with a small transverse momentum $Q_T \ll Q$.
Because both the total production rate and the production
characteristics, including the transverse momentum distribution, will
influence the search and analysis strategies for the Higgs,
a close study of higher-order corrections in the soft
and/or collinear limit is desirable.

Resummation techniques for Higgs production have been
established separately in the threshold~\cite{Higgsthreshold} and in the
recoil~\cite{HN,Higgsrecoil,Higgsresbos} cases, based on earlier
work~\cite{gs87,cat87,CSS} on the related case of Drell-Yan dimuon
production. A new analysis for threshold resummation, including the
recently calculated~\cite{NNLOsoftcolCdFG} next-to-next-to-leading
logarithmic (NNLL) coefficients $C^{(2)}$ and $D^{(2)}$,
can be found in~\cite{LesHouches02a,CdFGN}. The resummed Higgs-$Q_T$ 
distribution
has also recently been reinvestigated in~\cite{LesHouches02b,BQ,BCdFG}, using
the NNLL $B^{(2)}$ coefficient calculated in~\cite{dFG,CdFG},
and in \cite{BCdFG} matched to the Higgs cross section at
fixed $Q_T$ at NLO \cite{Kunszt}.
A complete understanding of soft gluon effects in differential
distributions also requires a study of the relation between the
two sets of corrections. A joint treatment of these corrections was
proposed in~\cite{Li,LSV}. It relies on a
refactorization of short-distance and long-distance physics at fixed
transverse momentum and energy~\cite{LSV}. In this formalism,
resummation of logarithmic corrections takes place in the impact 
parameter $b$ space~\cite{CSS}, Fourier conjugate to transverse 
momentum $Q_T$ space, and at the same time in Mellin-$N$ 
space~\cite{gs87,cat87}, conjugate to $z$ space. This guarantees 
simultaneous conservation of energy and transverse
momentum of the soft radiation. At present, the joint resummation 
formalism has been developed to next-to-leading logarithmic (NLL) accuracy.
A full phenomenological study of the joint resummation formalism,
as applied to electroweak Z production, was undertaken in~\cite{KSV}.

It may seem surprising that threshold resummation is relevant in
Higgs production at LHC energies, given that partonic threshold
for the production of a Higgs at rest involves partonic fractions
of the order of 10$^{-2}$, even smaller than for the Z boson at
the Tevatron, where the numerical effect of threshold resummation 
is rather small \cite{KSV}. Nevertheless, the results of 
\cite{NNLO,NNLOsoftcolCdFG,NNLOsoftcolHK} and 
\cite{Higgsthreshold,LesHouches02a,CdFGN} show that this is the case.  
Much of the difference between the Higgs and the Z boson can be attributed
to the difference between the color charge $C_A$ for the gluons that 
initiate
the former and $C_F$ for the quarks that initiate the latter.
This difference reflects itself in an increased sensitivity to
Sudakov logarithms associated with partonic threshold
for gluon-induced processes.
Gluon-induced processes, however, are also potentially sensitive
to radiation at low $x$. Indeed, the steep $x$-dependence of the 
gluon distribution functions at low $x$ enhances their sensitivity 
to threshold singularities.  By examining the
Higgs cross section at measured $Q_T$, we will find an intriguing
interplay between these two regions, which should have important
influence of the shape of the resummed distribution.

In any implementation of a resummation formula one has to address
the issue that the running strong coupling in the resummed expression
is probed at virtualities reaching down to $\Lambda_{\rm QCD}$ and 
below.
A number of  methods have been proposed to deal with this
singularity, while leaving intact the full resummed cross section
\cite{cspv,cmnt}, relying on the analytic
structure of the resummed cross section.
In our previous study of Z boson production,
\cite{KSV}, we made use of the minimal prescription \cite{cmnt}
for the Mellin-inverse,  and the prescription of
\cite{LSVPRL} for
the inverse Fourier transform, diverting it into the complex
$b$-plane according to the phase structure of
the integrand (see below).  In this way, we obtained a well-defined
resummed cross section, valid for all nonzero $Q_T$, which
described the $Q_T$ distribution of Z bosons produced at the Tevatron.
A complete fit requires a fitted non-perturbative parameter
in the form of a Gaussian in $b$ space \cite{Higgsresbos}.
In general, however, the effect of the nonperturbative parameter
is quite modest, as also found in Ref.\ \cite{qiuzhang}. Toward large
$Q_T$, a matching procedure between the fixed-order and the resummed 
results led to good agreement with the data there as well. The treatment
of the inverse $b$-space transform can be carried out as well
in pure $Q_T$ resummation as in joint resummation.  We can therefore
compare the two, to better understand the relationship between
threshold and transverse momentum resummation.

In this paper we study Higgs production via gluon-gluon fusion
for both $Q_T$
and  joint resummation. The leading order cross section is of order
$\as^2$ and is mediated via a heavy quark loop. Because the Higgs 
couples to fermions proportionally to their masses, we
make the  by-now standard simplification of
keeping only top quark loop contributions to the process.
We also treat the mass of the Higgs boson as much lighter
than twice the mass of the top quark.
Under these conditions, the Higgs-gluon
interaction can be described in the language of an effective
Lagrangian~\cite{NLO},  replacing the top quark loop by a simple 
effective $ggh$ vertex. It has been shown that the heavy top
quark approximation is accurate within a few percent in the case of NLO
calculations~\cite{Higgsthreshold} for actually a much wider range of
Higgs masses. Owing to similarities between the Drell-Yan production
mechanism and the gluon fusion mechanism for Higgs production (under the
above assumptions),  the joint formalism developed in~\cite{KSV}
can be fairly easily translated to the latter case. It turns out, 
however, that there are additional effects in the Higgs resummation case
that require a more detailed analysis, as we will show.
As suggested above, they are related to the gluon initial state 
for the reaction, resulting in amplified resummation effects through 
the basic replacement $C_F=4/3 \to C_A=3$ in the step from Drell-Yan to 
Higgs, and to the important role of $1/x$ terms in the 
splitting functions.

The joint resummation formalism for Higgs production in $N$ and $b$
space is described in section~2, along with its relations to threshold
and $Q_T$ resummations. In section~3, the method for inverting transforms 
is reviewed, along with the relation to more standard $Q_T$ resummations.
Section~4 presents our numerical results for the joint and pure $Q_T$ 
cases. In section~5 we discuss the total (i.e., $Q_T$-integrated) cross 
section and the possible role of terms not brought under control by joint
resummation. Finally, we draw conclusions.

\section{Joint resummation for Higgs production}

\subsection{The jointly-resummed cross section}

We start from the general formula for joint resummation~\cite{LSV,KSV},
applied to Higgs production:
\ba
\label{crsec}
       \frac{d\sigma_{AB}^{\rm res}}{dQ^2\,d^2 \vec Q_T}
       &=&   \pi \tau \sigma^h_{0} \delta(Q^2 -m_h^2)\,
H(\as(Q^2))
\int_{C_N}\, \frac{dN}{2\pi i} \,\tau^{-N}\;    \int \frac{d^2b}{(2\pi 
)^2} \,
e^{i{\vec{Q}_T}\cdot {\vec{b}}}\, \nonumber \\
&\times&    {\cal C}_{g/A}(Q,b,N,\mu,\mu_F )\;
        \exp\left[ \,E_{gg}^{\rm PT} (N,b,Q,\mu)\,\right] \;
        {\cal C}_{g/B}(Q,b,N,\mu,\mu_F) \; .
\ea
This is the same general form as for vector boson production,
but adapted to the Born cross section for Higgs production
in the heavy top mass limit,  $m_h \ll 2 m_t$.
The Higgs mass is $m_h$, and $\tau=Q^2/S$, where $Q$,
which is set equal to $m_h$ in the approximation
associated with threshold and $Q_T$ resummation,
sets the scale of the hard scattering.  As is characteristic
of joint resummation, the cross section is a double
inverse transform, in impact parameter $b$ and Mellin
moment $N$.  It resums logarithms in $b$ and $N$ to the level of
NLL.  Let us first summarize the content of the factorization
implicit in this expression.  

Each of the factors in Eq.\ (\ref{crsec}) is
associated with a specific range of perturbative dynamics involved
in the annihilation process.   These ranges are set by
the physical scales that enter the problem.  From soft to
hard these are, $\Lambda_{\rm QCD}$ at the scale of hadronic
binding, followed by scales introduced by the transforms: $Q/N$ and
$1/b$, followed by $m_h =Q$, $m_t$ and, finally $s$. We introduce
a function $\chi(bQ,N)$, to be specified in section~2.3., that
interpolates between $N$ in the threshold limit, $N\gg bQ$ and $bQ$ 
in the limit of large impact parameter, $bQ\gg N$, and is chosen in a way
that makes it possible to generate logarithms in $1/b$ and $Q/N$ at 
the level of next-to-leading logarithm in each.  At scales lower than 
$Q/\chi$, radiation is fully allowed, while beyond these scales, the
transforms in $N$ and $b$ constrain final-state radiation.
The resummations appropriate to these scales are
specified in terms of finite-order coefficients at each
logarithmic order.

\subsection{Coefficients and anomalous dimensions}

The shortest distances, with an off-shellness of order
$m_t$, are represented by
the perturbative function $H(\as)$. It collects the effects
of hard virtual corrections and reads to first
order~\cite{NLO,dFG}:
\be
\label{hfunc}
H(\as)=1+\frac{\as}{2\pi} \left(2 \pi^2+11\right)\; .
\ee
This is multiplied by the overall normalization constant
\ba
\sigma^h_0={\sqrt{2} G_F \as^2(m_h)  \over 576 \pi}
\ea
corresponding to the the heavy top
mass limit, $m_h \ll 2 m_t$, with $G_F$ the Fermi constant.

At the longest scales, the functions $\cal C$ represent evolution
between $\Lambda_{\rm QCD}$ and $Q/\chi(bQ,N)$. They are given
by
\be
\label{cpdf}
       {\mathcal C}_{g/H}(Q,b,N,\mu,\mu_F )
=   \sum_{j,k} C_{g/j}\left(N, \alpha_s(Q/\chi) \right)\,
{\cal E}_{jk} \left(N,Q/\chi,\mu_F\right) \,
                 f_{k/H}(N ,\mu_F) \; .
\ee
Here, the $f_{j/H}(N ,\mu_F)$ are the parton distribution
functions for hadron $H$ at any convenient factorization scale $\mu_F$.
The $C_{g/j}$ may be determined from the coefficients of the $\delta^2
(\vec Q_T)$ contributions in the order $\alpha_s$ correction to the
Higgs cross section \cite{CSS}, given our choice of $H^{(1)}$ in
Eq.\ (\ref{hfunc}).  They read
\ba
\label{ccoeff}
C_{g/g}\left( N,\as \right) &=& 1+
\frac{\as}{4\pi} \,\pi^2\; ,\\
C_{g/q}\left( N,\as\right) &=&\frac{\as}{2\pi} C_F \frac{1}{N+1}\; =\;
C_{g/\bar{q}}\left( N,\as\right)\;  .
\ea
The matrix ${\cal E} \left(N,Q/\chi,\mu_F\right)$ in Eq.\ (\ref{cpdf}) 
represents
the evolution of the parton densities from scale $\mu_F$ to scale
$Q/\chi$ to NLL accuracy in $\ln \chi$~\cite{KSV}.
In effect, the combinations of parton distributions, evolution factors
and matching coefficients follow the evolution of the
initial-state hadrons from the QCD scale up to the scale
at which the transforms suppress radiation.  

As indicated in Eq.~(\ref{cpdf}), we adopt the renormalization scale
$Q/\chi$ in the $C$-coefficients. This is guided by standard
$Q_T$-resummation, where it was argued~\cite{CSS} that the scale
should be proportional to 
$1/b$. The choice of the scale in the $C$-coefficients
actually only affects NNLL terms in the resummation.

We note that in the inclusive limit $b=0$ the coefficients
$C_{g/j}$ and $H$ do not quite reduce to the corresponding factor 
in threshold resummation. First, there is no 
analog of the nondiagonal $C_{g/q}$ and $C_{g/\bar q}$ terms in 
the pure threshold case, since they are not associated with logarithmic 
behavior at large $N$. Second, the scale of $\alpha_s$ in
$C_{g/j}$ reduces to $Q/\bar N$ rather than $Q$ for
$b=0$. Finally, for the diagonal coefficient in our 
formulas above the combined effect at ${\cal O}(\as(Q^2))$ is 
$H\times C_{g/g} \times C_{g/g} \approx 1 + \frac{\as}{2\pi}
\left(11 + 3 \pi^2 \right)$, whereas the corresponding coefficient 
in the threshold case is \cite{CdFGN} $1 + \frac{\as}{2\pi}
\left(11 + 4 \pi^2 \right)$. Such differences are present as 
well in the case of Z boson production through
quark annihilation \cite{KSV}.  A development of the joint 
formalism to NNLL in both the threshold and small-$Q_T$
limits would account for them.  

We recall that organizing the ${\cal C}$ functions in the
form of Eq.\ (\ref{cpdf}) in Mellin-$N$ moment space
has a great advantage: it enables us to explicitly separate
the evolution of the parton densities between the scales $\mu_F$ and 
$Q/\chi$, embodied by the matrix ${\cal E}$  \cite{KSV}. In this way, 
we can avoid the problem normally faced in $Q_T$
resummation that one needs to call the parton densities at scales
far below their range of validity, so that some sort of ``freezing''
(or related prescription) for handling them is required.
Our matrix ${\cal E}$, on the other hand, may be expanded
to LL and NLL accuracy, consistent with all our approximations,
as shown in ref.~\cite{KSV}. We then only need the parton distribution 
functions at the ``large'' scale $\mu_F \sim Q$, whereas normally in 
$Q_T$ resummation the product $\sum_k {\cal E}_{jk} \left(N,1/b,
\mu_F\right) \,f_{k/H}(N ,\mu_F)$ is identified with $f_{j/H}(N ,1/b)$.
We also recall that the moment variable $N$ and, as will be discussed 
below, also the impact parameter $b$ are in general complex-valued in 
our approach, so that it is even more desirable to separate the complex 
scale $Q/\chi$ from that in the parton densities. In practical 
applications, one usually has available only parton distributions in 
$x$-space. This obstacle can be overcome in various ways. We have 
described one practical approach in Appendix~B of~\cite{KSV}. A very 
simple other method is to fit the parton
distributions at scale $\mu_F$ in the $x$-region of relevance
by a simple functional form that allows to take Mellin-moments
analytically~\cite{ddfpriv}.  We have used both methods and found
excellent agreement between the corresponding numerical results.

The exponential $E_{gg}^{\rm PT} (N,b,Q,\mu)$ in Eq.\  (\ref{crsec}) 
is the Sudakov factor for Higgs production in joint resummation.  
It summarizes QCD dynamics in the transverse momentum range,
$Q/\chi \le k_T \le Q$, where real radiation
is kinematically constrained and virtual corrections produce
an exponentiating double-logarithmic suppression.  We
may think of it as interpolating between the scale of the
partonic coefficients $\cal C$ and the short-distance
annihilation process. $E_{gg}^{\rm PT} (N,b,Q,\mu)$ is equivalent, 
with trivial changes in color factors, to the corresponding exponent 
in Z production \cite{KSV}, and is given by
\be
\label{sudakov}
E_{gg}^{\rm PT} (N,b,Q,\mu)=
-\int_{Q^2/\chi^2}^{Q^2} {d k_T^2 \over k_T^2} \;
\left[ A_g(\as(k_T))\,
\ln\left( {Q^2 \over k_T^2} \right) + B_g(\as(k_T))\right]  \; ,
\ee
accurate to NLL level.
Dependence on the renormalization scale $\mu$ is implicit in
Eq.~(\ref{sudakov})
through the expansion of $\as(k_T)$ in powers of $\as(\mu)$ and
logarithms of $k_T/\mu$.

The anomalous dimensions in Eq.\ (\ref{sudakov}) are
perturbative series in $\as$, $A_g(\as)=\sum_{i=0}^{\infty} \left({\as 
\over \pi}\right)^i A_g^{(i)}$, and similarly for $B_g$.
By examining the fixed-order cross section at NLO~\cite{NLO}
 the following familiar values necessary for NLL resummation can be
derived~\cite{dFG,KT},
\ba \label{abcoeffs}
A_g^{(1)} &=&  C_A\; , \qquad\qquad
B_g^{(1)}\;=\;-\frac{1}{6} \left( 11 C_A - 4 T_R N_F 
\right)\;,\nonumber \\
A_g^{(2)} &=& \frac{C_A}{2} \left[
C_A \left( \frac{67}{18}-\frac{\pi^2}{6} \right) -\frac{10}{9}T_R
N_F\right]\; ,
\ea
where $C_A=3$, $C_F=4/3$, $T_R=1/2$, and $N_F$ is the number of
flavors. NLL accuracy requires using $A_g^{(1)},\ B_g^{(1)}$ and 
$A_g^{(2)}$
in Eq.~(\ref{sudakov}).

 From an NNLO~\cite{NNLO} calculation one can also
determine a second-order coefficient for the function $B(\alpha_s)$.
This coefficient contributes to the cross section at the level of NNLL.
Given our choices for $H^{(1)}$  and hence for $C^{(1)}$,
and assuming the NLL expression for the Sudakov exponent~(\ref{sudakov}), 
this coefficient  is uniquely determined to be
\ba \label{Btwo}
B_g^{(2)}&=&
C_A^2\left(-{4 \over 3}+{11 \over 36}\pi^2 -{3 \over2}\zeta_3\right)
+ {1 \over 2}C_F T_R N_F + C_A N_F T_R
\left({2 \over 3} - {\pi^2 \over 9} \right)
\; .
\ea
Even though the contributions associated with the 
coefficients $H^{(1)}$, $C^{(1)}$ and $B_g^{(2)}$
are all only of NNLL order, we will include them
in our study. It was found in previous studies on $Q_T$ 
resummation for Higgs production that the combined
contribution of these coefficients is actually
numerically rather significant, partly due to the size of $C_A$
relative to $C_F$.

As was shown in ref.~\cite{CdFG}, there is actually some
freedom in how one apportions contributions to $H^{(1)}$, $C^{(1)}$ 
and $B_g^{(2)}$ individually. A choice for one coefficient
uniquely determines the others, but it is possible
to shift terms among them, resulting in different 
but formally equivalent schemes for the resummation. 
We believe that our choice in Eq.~(\ref{hfunc}) is 
well motivated physically, since with this choice the function 
$H$ contains all genuinely hard short-distance effects at scales 
above $Q$. The function $B_g^{(2)}$, on the other hand,
enters with dependence on scales much smaller than $Q$, as
can be seen from Eq.~(\ref{sudakov}), so excluding hard
contributions from $B_g^{(2)}$ seems natural.
We have nevertheless also investigated an alternative choice,
for which all terms in $H$ are absorbed into the ${\cal C}$ functions,
so that $H=1$, with $B_g^{(2)}$ adapted accordingly. This 
scheme has been more customary in previous studies. We found that 
our numerical results presented below do not change significantly if this
``resummation scheme''~\cite{CdFG} is used. We note that for our
choice of $H^{(1)}$, unlike for the $H=1$ scheme,
the corresponding numerical value for $B_g^{(2)}$ is very small.  
A more detailed discussion of the interrelations between $H^{(1)}$, 
$C^{(1)}$ and  $B_g^{(2)}$ may be found in~\cite{CdFG}.

There are other NNLL contributions in the exponent, besides 
those associated with $H^{(1)}$, $C^{(1)}$ and $B_g^{(2)}$. 
The third-order coefficient $A_g^{(3)}$ also contributes at NNLL. 
Its value is so far known numerically \cite{vogta3}, with its 
contributions proportional to $N_F^2$ and $N_F$ determined 
analytically~\cite{gracey,bergera3,vogt3loop}. We have checked that when we 
include $A_g^{(3)}$ at the level of its currently known approximation
the impact on our numerical results is insignificant.

We also note that in the context of joint resummation 
the form of Eq.\ (\ref{sudakov}) for the Sudakov exponent is 
itself only valid to NLL. The original jointly resummed Sudakov
exponent given in~\cite{LSV} is, in fact, far more general than
NNL and will correctly reproduce logarithms of any order
for threshold and $Q_T$ resummation in the limits of large $N$ and $b$,
respectively. For large $b$ and fixed $N$, the structure in 
(\ref{sudakov}) is correct to all logarithmic order, as is 
known from standard $Q_T$ resummation, where $\chi \sim b$. 
In the opposite 
(threshold) limit of $N$ large and $b$ finite, we should in 
general introduce an additional function that reflects coherent 
soft radiation at wide angles, called $g_3$ in \cite{gs87} and 
$D$ in \cite{vogt3loop,vogtnnll}. For the purposes of this study, we 
follow standard $Q_T$ resummation as closely as possible and therefore
use the form (\ref{sudakov}) also at NNLL. We emphasize at this point
that a full investigation of NNLL effects in the joint resummation 
formalism would be very desirable. Our taking into account of $H^{(1)}$, 
$C^{(1)}$ and $B_g^{(2)}$ can therefore be only an
approximation for the full NNLL effects. We defer a full NNLL analysis
to a future publication.

In the next subsection we discuss the Sudakov exponent in more 
detail.

\subsection{The Sudakov exponent and the choice of $\chi$}

The exponent $E_{gg}^{\rm PT} (N,b,Q,\mu)$
in Eq.\ (\ref{sudakov}) isolates the
Sudakov suppression in the logarithm of the full NLL eikonal cross 
section in $N$- and $b$-space \cite{LSV},
\ba
E_{gg}^{\rm (eik)}(N,b,Q,\mu_F)  &=&2\,
\int_0^{Q^2} {d k_T^2\over k_T^2}\; \left\{\ 
A_g\left(\as(k_T)\right)\; \left[\, J_0 \left( b k_T \right) \;
K_0\left({2Nk_T\over Q} \right) + \ln\left({\bar N k_T\over
Q}\right)\, \right]\,
\right\}
\nonumber\\
&\ & \hspace{-3mm}
-2\, \ln \bar N\ \int_{\mu_F^2}^{Q^2} {d k_T^2\over k_T^2}
A_g\left(\as(k_T)\right)\, ,
\label{Eelaborate}
\ea
where here and below, we denote
\ba
\label{nbdefs}
\bar{N} = N{\rm e^{\gamma_E}}\, ,
\ea
with $\gamma_E$ the Euler constant.
As in Ref.\ \cite{KSV}, we rewrite Eq.\ (\ref{Eelaborate}) in a form
that is accurate to next-to-leading logarithm in both $b$ and
$N$,
\ba
E_{gg}^{\rm eik} (N,b,Q,\mu,\mu_F) &=&
-\int_{Q^2/\chi^2}^{Q^2} {d k_T^2 \over k_T^2} \;
\left[ A_g(\as(k_T))\,
\ln\left( {Q^2 \over k_T^2} \right) + B_g(\as(k_T))\right] \nonumber \\
&+&\int_{\mu_F^2}^{Q^2/\chi^2} {d k_T^2 \over k_T^2}
\Big[ -2 A_g(\as(k_T)) \ln\left( \bar{N}\right)
- B_g(\as(k_T)) \Big] \; .
\label{Elog2}
\ea
Here we have added and subtracted a term with $B_g$, the term 
constant in $N$ in the DGLAP splitting functions. To NLL, $B_g$
coincides with the function defined in the previous section.
We have also neglected logarithms of the factorization scale divided 
by $Q$. The Sudakov exponent is the first term in this expression. 
We will reinterpret the second term as the large-$N$ limit of
the evolution of parton distributions in the next section.

The function $\chi(bQ,N)$ in Eq.~(\ref{Eelaborate}) organizes logarithms 
of $N$ and $b$ in joint resummation~\cite{KSV} and is given by
\be
\label{chinew}
\chi(bQ,N)=\bar{b} + \frac{\bar{N}}{1+\eta{\bar{b} \over 
\bar{N}}}\; ,
\ee
where we define
\ba
\bar{b}\equiv b Q {\rm e^{\gamma_E}}/2 \; .
\ea
This form of $\chi(N,b)$ is chosen to reproduce threshold logarithms 
at NLL for large $N$ at fixed $b$, and impact parameter
logarithms at large $b$ with $N$ fixed.   The functional
form ensures that corrections at low $Q_T$ (hence, at large $b$)
avoid unphysical singularities suppressed by $1/Q_T$
compared to leading terms: as discussed in~\cite{KSV}, any 
nonzero value for the parameter $\eta$ in Eq.\ (\ref{chinew}) 
eliminates corrections of order $b^{-1}$ in the large-$b$ limit,
relative to leading behavior.  It therefore does not introduce
spurious logarithmic singularities in $Q_T d\sigma/dQ^2dQ_T^2$
as $Q_T\rightarrow 0$. To NLL, the cross section is independent of 
$\eta$, as is the $Q_T$-integrated cross section, found by setting 
$b$ to zero. The shape of the cross section depends to some extent on
the precise value of $\eta$ that we choose, through subleading
terms. For very large values of $\eta$, $\chi$ approaches its value 
in ``pure recoil" $b$-space resummation, $\chi=\bar b$, although the 
integrated cross section is then ill-defined.

In the following, we choose the value of $\eta=1$.
We motivate this choice by comparing the expansions of the logarithm of 
$\chi$ in terms of $\ln \bar b$ at fixed $\bar N$, and of $\ln\bar N$
at fixed $\bar b$. For $\eta=1$ these expansions give rise to ``power-like"
corrections of the form $(\bar N/\bar b )^2$ in the former case, 
and $(\bar b/\bar N)^2$ in the latter, with no linear $\bar{b}/\bar{N}$
corrections. We readily verify that the full NLL exponent 
(\ref{Eelaborate}) also has no corrections linear in $b/N$ for $b/N 
\rightarrow 0$. The choice $\eta=1$ 
treats the remaining two sets of corrections, 
$(\bar N/\bar b )^2$ and $(\bar b/\bar N)^2$, symmetrically.  
To estimate the sensitivity to $\eta$, we will also exhibit 
results for $\eta=1/2$ and $\eta=2$ below. In Ref.~\cite{KSV}, 
on the other hand, where we used the framework of joint resummation 
to analyse Z production at the Tevatron, a choice of $\eta=1/4$ 
was made. It led to a relatively simple structure of singularities 
in the complex $b$ plane. We checked that the $\eta$ dependence 
of the $Q_T$ distribution for Z bosons was insignificant~\cite{KSV}. 
This is, however, not necessarily the case for Higgs production at the LHC.
The treatment of the inverse transforms (i.e. contours in the
complex $N$ and $b$ spaces) for differing values of
$\eta$ is discussed in section 3.

The expansion of $E_{gg}^{\rm PT}$ into leading, next-to-leading,
and next-to-next-to-leading logarithms gives the form~\cite{LSV,KSV}
\begin{equation}
\label{expdef}
E_{g g}^{\rm PT,\,NNLL} (N,b,Q,\mu) = \frac{2}{\alpha_s (\mu)}\,
h_g^{(0)} (\beta) +
2\,h_g^{(1)} (\beta,Q,\mu) + 2\,\alpha_s (\mu)\, h_g^{(2)}
(\beta,Q,\mu)  \;  ,
\end{equation}
where each function $h_g^{(i)}$ is a power series in
\begin{equation}
\beta = b_0\, \alpha_s (\mu)
\ln \left( \chi \right) \, ,
\label{varsdef}
\end{equation}
with
\ba \label{b0def}
b_0 &=& \frac{11 C_A - 4 T_R N_F}{12 \pi}\; .
\end{eqnarray}
The leading- and next-to-leading logarithmic functions
$h_g^{(0)}$ and $h_g^{(1)}$ have already been presented 
in~\cite{LSV,KSV}.
For convenience, we provide them again in Appendix~A.
As stated above, we do not attempt a full
NNLL study of the resummed cross section, but we do wish to
include those NNLL terms that are known to produce numerically
significant effects. At the level of the Sudakov exponent,
this term is given by the contribution involving $B_g^{(2)}$
and gives rise to the function $h_g^{(2)}$:
\begin{eqnarray}
h_g^{(2)} (\beta,Q,\mu)&=& -{B_g^{(2)} \over \pi^2 b_0}{\beta \over
(1- 2 \beta)} + \ldots \; ,
\label{hsub2adef}
\end{eqnarray}
where the ellipses denote further, neglected,
terms proportional to $A_g^{(1)}, B_g^{(1)}, A_g^{(2)}$, as well 
as to the third-order coefficient $A_g^{(3)}$. As mentioned above, we 
have checked that those terms among the
neglected terms that are fully known lead to numerically insignificant
effects.  

\subsection{Relation to threshold resummation}

In the resummed cross section, Eq.\ (\ref{crsec}),
the partonic functions ${\mathcal C}_{g/H}(Q,b,N,\mu,\mu_F )$,
Eq.\ (\ref{cpdf}), incorporate
the parton distribution functions $f_{j/H}(N ,\mu_F)$, the evolution
matrices ${\cal E}_{jk} \left(N,Q/\chi,\mu_F\right)$, and the standard 
$C$
coefficients coming from matching the fixed order to the resummed 
result.

Within a strict threshold approximation, only parton-diagonal 
contributions to evolution are taken into account, which 
for Higgs production means the sole use of the large-$N$ limit
of the $gg$ anomalous dimension.
The matrix ${\cal E} \left(N,Q/\chi,\mu_F\right)$
then only has an entry in the $gg$ position and reads
\begin{equation} \label{gevol}
{\cal E}(N,Q/\chi,\mu_F)=\exp\left[ \frac{-2A_g^{(1)}\ln\bar{N} -
B_g^{(1)}}{2\pi b_0} \, s(\beta) \right] \;
\left(  \begin{array}{cc}
0& 0\\
0& 1
\end{array}
\right) \; ,
\end{equation}
where~\cite{KSV}
\begin{equation}
\label{sbeta}
s(\beta)\equiv \ln\left( \frac{\as(\mu_F)}{\as(Q/\chi)} \right)
=\ln (1-2 \beta) \;+ \;\mbox{NLL}\; .
\end{equation}
In Eq.~(\ref{gevol}) we have for the sake of brevity
only written down the LO/LL
part of the evolution;  extension to NLO/NLL is straightforward and
included in our analysis. The ${\cal E}$ in Eq.~(\ref{gevol}) may then 
be
combined with the Sudakov exponent in~(\ref{expdef}) at $\beta=
b_0 \alpha_s \ln \bar{N}$
(i.e., $b=0$) to reproduce the familiar exponent of
Higgs threshold resummation~\cite{Higgsthreshold,LesHouches02a,CdFGN}.

However, as we discussed in~\cite{KSV}, one can improve the above
near-threshold approximation and use the full NLO singlet evolution
matrix in ${\cal E}$, with full $N$-dependence for the anomalous 
dimensions.
This amounts to the replacement
\be
-2 A_g(\as) \ln\left( \bar{N}\right) - B_g(\as) \; \longrightarrow \;
\gamma_N (\as) \; ,
\label{repl}
\ee
or equivalently, in terms of parton distribution functions,
\be
\delta_{jg}\, \exp\left[ \frac{-2A_g^{(1)}\ln\bar{N} -
B_g^{(1)}}{2\pi b_0} \, s(\beta) \right] \,  f_{g/H}(N ,\mu_F)\;
\longrightarrow \;
{\cal E}_{jk} \left(N,Q/\chi,\mu_F\right) \,
                 f_{k/H}(N ,\mu_F)\, .
\label{pdfrepl}
\ee
In this way, ${\cal E}$, and hence the functions $\cal C$
in Eq.\ (\ref{crsec}), obtains a full matrix structure and becomes
an ordered exponential.
On the other hand, the standard threshold resummation formula
is recovered from Eq.\ (\ref{crsec})
by integrating over $\vec Q_T$ to set $b=0$, and then
undoing the replacements of Eqs.\ (\ref{repl}) and (\ref{pdfrepl}).

One of the assets of the joint resummation
of Eq.\ (\ref{crsec}), incorporating (\ref{pdfrepl}), is to resum
collinear, but non-soft, terms of the form $\as^n \ln^{2n-1} (\chi)/N$
to all orders~\cite{KSV}. Such terms are obviously subleading with 
respect to the threshold approximation, but may have some relevance for
phenomenology. They were discussed 
in~\cite{Higgsthreshold,LesHouches02a,CdFGN}
in the context of threshold resummation, where they occur as
$\as^n \ln^{2n-1} (\bar{N})/N$. In this formalism they are brought 
under control automatically. In our analysis of Higgs production, 
we actually find that other formally subleading terms can also be 
rather important. We will discuss this issue in section 5 below.

\section{Inverse transforms and matching}

We have already mentioned in the introduction the importance of suitably
defining the inverse Mellin and Fourier transforms.
We follow here closely our previous phenomenological study for 
electroweak
annihilation~\cite{KSV}, where all details may be found. Let us
briefly recall the main ingredients.

The contour for the inverse Mellin transform is chosen
according to the ``minimal prescription'' of~\cite{cmnt}.
It is parameterized as
\be
\label{cont}
N = C+ z {\rm e}^{\pm i \phi} \; ,
\ee
where the upper (lower) sign applies to the upper (lower)
branch of the contour, with $0\leq z\leq \infty$ ($\infty\geq z\geq 0$).
The constant $C$ is required to lie to
the right of the rightmost singularity of the parton distribution 
functions.
At any finite order in perturbation theory, all values of $C>0$, and all
angles $\pi>\phi>\pi/2$ are equivalent.  In the resummed cross section,
however, the singularity at $\chi=\rho_L=\exp[1/2b_0\as(\mu)]$ 
introduces
an ambiguity in the transform, which may be resolved by choosing
$C<\rho_L$~\cite{cmnt}.

\begin{figure}[h]
\begin{center}
\vspace*{2mm}
\epsfig{file=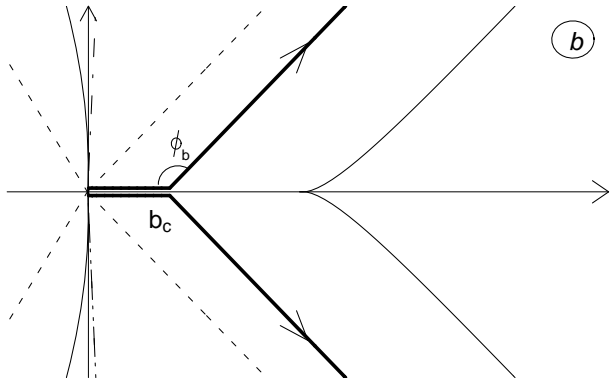,width=8cm}
\caption{Choice of contour for $b$ integration (thick solid lines) 
for $\eta=1$. The straight sections of the contour from $0$ to $b_c$ 
are to be interpreted as on the positive real axis. The remaining curves 
represent
lines of singularity discussed in~\protect{\cite{KSV}}.}
\label{bcont}
\end{center}
\end{figure}

The inverse Fourier integral is also defined as a contour
integral~\cite{KSV,LSVPRL}. It is performed using the identity
\be
\label{bint}
\int d^2 b \;e^{i\vec{q}\cdot {\vec{b}}}\,f(b)\;=\;  2 \pi\,
\int_0^{\infty} \, db\,b\, \,J_0(bq) \,f(b) \;=\; \pi\,
\int_0^\infty db\, b\, \left[\, h_1(bq,v) + h_2(bq,v)\,\right]\,f(b) \, 
,
\ee
and employing Cauchy's theorem to deform the integration over real $b$
into a contour in the complex $b$ plane~\cite{LSVPRL,KSV},
as shown in Fig.~\ref{bcont}. Here the auxiliary functions $h_{1,2}$ 
\cite{LSVPRL} are related to Hankel functions.
They distinguish between the positive and negative phases in
Eq.~(\ref{bint}). The $b$ integral can thus be written as a sum of two
contour integrals, of the integrand with $h_1$ ($h_2$) along a contour
in the upper (lower) half of the $b$ plane. The precise form of the 
contours becomes unimportant as long as the contours do not run into 
the Landau pole or singularities associated with the particular 
form~(\ref{chinew}) of the function $\chi$. One also needs to make 
sure that the lines of singularities do not cross, which would make 
it impossible to draw a fixed contour without running into a line. This can 
be ensured by choosing the opening angle of the $N$ contour to fulfill 
the condition $\pi/2<\phi<\pi- \tan^{-1}\left( \sqrt{|4\eta-1|} \right)$.

This choice of contours in complex transform space is
completely equivalent to the original form, Eq.\ (\ref{bint}) when
the exponent is evaluated to finite order in perturbation theory.
In the presence of the Landau pole arising in the resummed formula,
it is a natural extension of the $N$-space contour redefinition
above~\cite{cmnt}, using a generalized ``minimal'' exponent~(\ref{expdef}). 
We emphasize that joint resummation with its contour
integration method  provides an  alternative to
the standard $b$ space resummation.  Joint resummation has built-in
a \prt treatment of large $b$ values, eliminating the need for a
$b_*$ \cite{CSS} prescription for the exponent, or for a freezing of the
scale of parton distributions at large $b$ or low $Q_T$. In this way,
we can derive entirely perturbative cross sections for Higgs 
production. We note that this definition of the $b$-space contour 
was also adopted in the recent study \cite{BCdFG}, while in 
\cite{qiuzhang} an extrapolation of the perturbative exponent to 
large $b$ is used to define the inverse transform. 

At large values $Q_T\sim m_h$, we need to match the resummed
cross section to fixed-order perturbation theory. This is
achieved in the following simple way~\cite{BCdFG,KSV}:
\begin{equation}
{d \sigma \over d Q^2 d Q_T^2} = {d \sigma^{\rm res} \over d Q^2 d 
Q_T^2}
-  {d\sigma^{\rm exp(k)} \over d Q^2 d Q_T^2} +
{d \sigma^{\rm fixed(k)} \over d Q^2 d Q_T^2} \,,
\label{joint:match}
\end{equation}
where $d \sigma^{\rm res}/d Q^2 d Q_T^2$ is given in Eq.~(\ref{crsec}) 
and
$d\sigma^{\rm exp(k)}/d Q^2 d Q_T^2$ denotes the terms resulting from
the expansion of the resummed expression in powers of
$\as(\mu)$ up to the order ${\rm k}$ at which the fixed-order cross section
$d \sigma^{\rm fixed(k)} /d Q^2 d Q_T^2$  is taken. The last two terms 
on
the right-hand-side of Eq.~(\ref{joint:match}) should cancel each other 
at
small $Q_T$, since order by order the terms generated by the resummed
formula must reproduce the singular behavior of the cross section
at small $Q_T$.  In our study, we do the matching using ${\rm k}=1$ 
(see~\cite{HN}) in Eq.~(\ref{joint:match}).

\section{Predictions for the resummed Higgs cross section}

For our calculations of the Higgs transverse momentum distribution at
the LHC, obtained in the framework of joint resummation,
we choose $m_h= 125$~GeV and the factorization and renormalization
scales $\mu=\mu_F=Q=m_h$. We use the CTEQ5M~\cite{cteq5m}
parton distribution function
parameterizations. The contour parameters are $\phi=17/32 \pi$ and
$\phi_b=\phi+\tan^{-1} \sqrt{|4\eta-1|}$, with $\eta=1$.

\begin{figure}[h]
\begin{center}
\epsfig{file=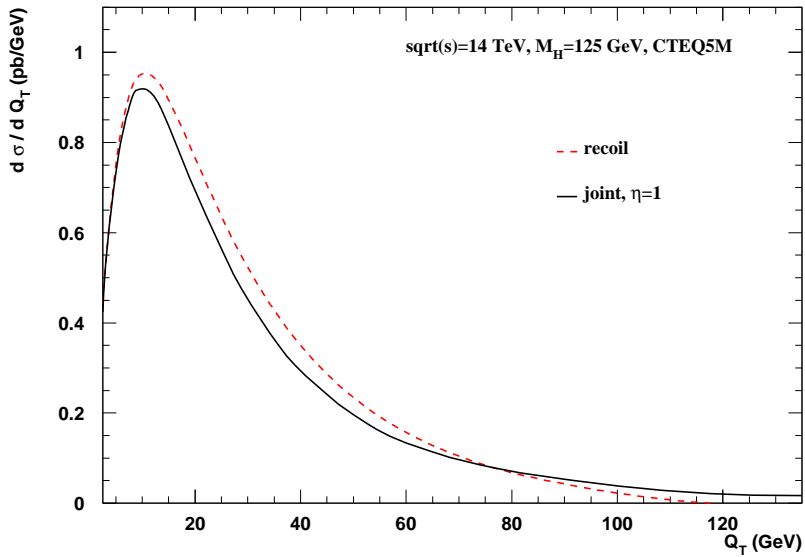,width=12cm}
\caption{Transverse momentum distribution for Higgs production at the LHC in
the framework of joint resummation and of ``pure-$Q_T$'' resummation.}
\label{higgs}
\end{center}
\end{figure}

Fig.~\ref{higgs} shows our result for $d\sigma/dQ_T$ as a
function of $Q_T$. The solid curve shows the result obtained
for the formulas given in Sec.~2. For comparison, we also
display the result for ``pure-$Q_T$'' resummation, i.e., for
the case when we do not take into account threshold resummation
effects, $\chi=\bar{b}$ (dashed line).
We find that the corresponding result is
somewhat higher and broader than our jointly resummed curve.
We can compare this curve to previous analyses in the
literature~\cite{Higgsresbos,LesHouches02b,BQ}. It turns out
that our ``pure-$Q_T$'' resummed result is about 30\% higher
(in the peak of the distribution) than the one of the
{\sc Resbos}~\cite{LesHouches02b} code, and about 20\% higher
than the result of~\cite{BQ}. On the other hand, our distribution 
is lower by about 15\% than the NLL result of~\cite{BCdFG}, 
matched with the NLO predictions. These differences appear to 
be due in part to different treatments of the large-$b$ region in
the various formalisms, and to different organizations
of the coefficients.

The jointly resummed distribution in Fig.~\ref{higgs} was obtained after 
adding a small non-perturbative term of the form 
$-g\,b^2$ to the exponent. As discussed
in~\cite{KSV}, it is an advantage of joint resummation that no
non-perturbative contribution is required in order to obtain
results for any nonzero $Q_T$.  The reason for introducing a 
non-perturbative component here is purely technical. Our choice of 
$\eta=1$ necessitates small values of angle $\phi$, making it time consuming
to achieve numerical stability at low $Q_T$. The value of the parameter $g$ 
was adopted from the study~\cite{KS3}, where an analysis of Upsilon
hadroproduction was performed as a means of studying the non-perturbative 
contributions in processes with two gluons in the initial state.
We use the value $g=1.67$~GeV$^2$ corresponding to the lower end of
a range of values extracted from data. In any case, on the basis of
previous studies of the dependence of $b$-space resummation
for Higgs production on the non-perturbative 
contribution~\cite{BQ,KS3,BCS}, we expect the dependence on the parameter 
$g$ to be weak for values of $Q_T$ around or above the peak of the 
cross section.  In effect, a nonzero value of $g$ simply enables 
us to extrapolate the cross section to $Q_T=0$.

\begin{figure}[h]
\begin{center}
\epsfig{file=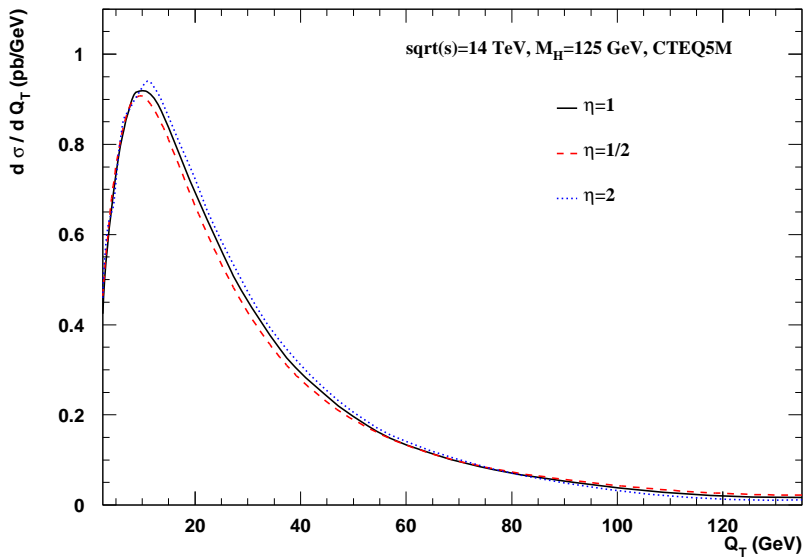,width=12cm}
\caption{Dependence of the jointly resummed transverse momentum 
distribution for Higgs
production at the LHC on the value of the parameter $\eta$.}
\label{etadep}
\end{center}
\end{figure}

In section~2 we introduced the parameter $\eta$ in the definition of the
function $\chi$, Eq.~(\ref{chinew}), and as discussed above chose $\eta=1$. To
investigate the dependence of the results on the value of this parameter, we
computed the $Q_T$ distributions for $\eta=1/2$ and $\eta=2$.
As can be seen from Fig.~\ref{etadep} the numerical sensitivity of the 
results to the value of $\eta$ is rather small in this range.

\section{Joint resummation and the role of non-threshold \\ effects}

In a sense, Fig.\ \ref{higgs} completes our study, since it
predicts a cross section that remains to be measured.
Nevertheless, a number of cross checks in
joint resummation suggest that Fig.\ \ref{higgs} is not
the final word, and that to get a full picture of
the Higgs cross section in $Q_T$ it may be desirable to
extend the formalism to include the contributions that
give rise to dominant small-$x$ behavior.

In this section, we compare the jointly resummed cross
section in more detail to existing fixed-order and threshold-resummed
calculations.

\subsection{NLO expansion in $Q_T$ space}

\begin{figure}[h!]
\begin{center}
\epsfig{file=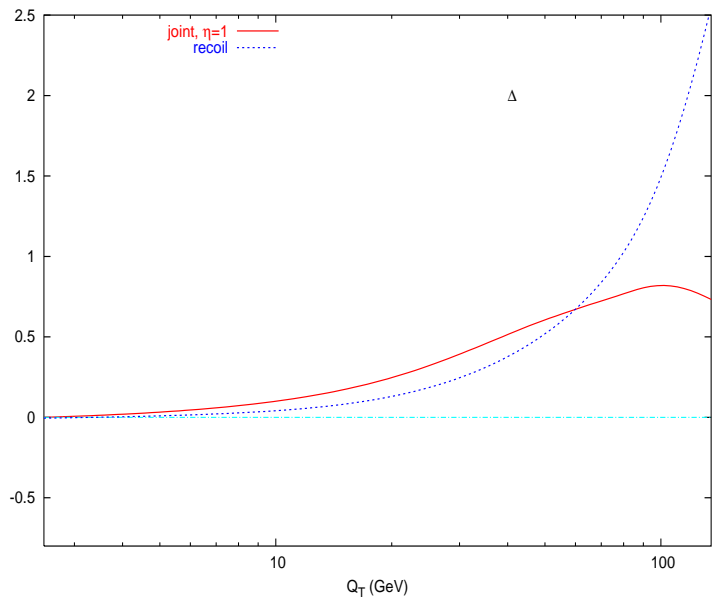,width=8cm,height=10cm,angle=270}
\caption{Fractional deviation $\Delta$ (as defined in Eq.~(\ref{frdev}))
between the ``exact'' $\CO (\as)$ result and the $\CO (\as)$
expansion of the jointly and the pure-$Q_T$ resummed cross sections.
\label{delta}
}
\end{center}
\end{figure}

Fig.~\ref{delta} shows the difference
$d \sigma^{\rm fixed(1)} /d Q_T - d\sigma^{\rm exp(1)} / d Q_T$ and
the fractional deviation
\begin{equation}
\Delta\equiv\left[
{d \sigma^{\rm fixed(1)} \over d Q_T} -
{d\sigma^{\rm exp(1)} \over d Q_T}\right] \;/\; {
d \sigma^{\rm fixed(1)} \over d Q_T} \,,
\label{frdev}
\end{equation}
using $m_h= 125$~GeV, scales $\mu=\mu_F=Q=m_h$, and the
CTEQ5M~\cite{cteq5m} parton distribution functions.
We find very good agreement in the region $Q_T<10$ GeV where
resummation is necessary. Beyond $10$ GeV, the agreement is naturally
less exact but still reasonable. This is the region where matching
to finite order is appropriate, as expressed by Eq.~(\ref{joint:match}).
From the figure we see that the `recoil' approximation, $\chi\rightarrow 
\bar b$, is slightly more accurate at low $Q_T$, while the joint form remains 
stable at larger values of transverse momentum.  The somewhat lower value 
of the expanded joint cross section is due in part to the lower scale at which
the gluon distribution is evaluated compared to pure $Q_T$-resummation 
($Q/\chi$ as opposed to $1/b$).  At larger values, the stability of the 
jointly resummed cross section makes it easier to match to fixed order 
(a property shared with the $Q_T$-resummed treatment in Ref.~\cite{BCdFG}).

\subsection{Total cross section and subleading terms in $N$ space}

When integrated over $Q_T$, the jointly resummed distribution formally
reduces to the threshold resummed cross section in the large-$N$ limit. 
When we numerically compare 
the two formalisms, however, we find that the total cross section in 
joint resummation is around ten percent lower than pure threshold 
resummation at NLL: after matching to NLO we find the NLL 
threshold-resummed cross section to be $39.4$ pb (consistent with 
results shown in~\cite{CdFGN}), whereas our jointly resummed 
cross section integrates to $35.0$ pb. This slight disagreement 
implies that terms that are subleading with regard to the threshold 
approximation play a fairly important role. Since our formalism
has been designed to automatically reproduce all behavior 
singular at $Q_T \to 0$, we are led to associate the ``missing
area'' with contributions to the Higgs cross section at rather high 
$Q_T$. 

It is instructive to investigate these points in more detail. 
To this end, we expand the jointly resummed, and $Q_T$-integrated, 
cross section to fixed order $\as$ (not counting the overall
$\as^2$ of the cross section), and compare it to
the ``exact''  expressions for the Higgs production
cross section at this order. For simplicity, we set the renormalization
and factorization scales to $m_h$. Expansion of Eq.~(\ref{crsec}) to 
$\CO(\as)$ gives for the partonic cross sections in the $gg$ and
$gq$ scattering channels:
\begin{eqnarray} \label{sec1}
\hat{\sigma}^{gg}&=& \sigma_g^{(0)} \,\frac{\as}{2\pi}\,\left\{
-4 C_A\ln^2\bar{N}+8 \pi b_0 \ln\bar{N}
+11+3\pi^2\right.
\nonumber \\
&&
\hspace*{1cm}-2 \ln\bar{N}
\left.
\left[ \frac{4C_A}{N(N-1)}+\frac{4C_A}{(N+1)(N+2)}-4 C_A S_1(N)
+4 \pi b_0\right] \right\}
\quad , \nonumber  \\
\hat{\sigma}^{gq}&=&  \sigma_g^{(0)} \,\frac{\as}{2\pi}\,
C_F\,\left\{ -2 \ln\bar{N}\frac{N^2+N+2}{N(N+1)(N-1)} +
\frac{1}{N+1}\right\}\; ,
\end{eqnarray}
where $S_1(N)=\sum_{j=1}^N j^{-1}=\psi(N+1)+\gamma_E$, with $\psi$
the digamma function. Since we are interested in the near-threshold
region, we can expand this further to the large-$N$ limit.
Using $\psi(N+1)= \ln N+1/(2N)+{\cal O}(1/N^2)$, we find:
\begin{eqnarray} \label{largen1}
\hat{\sigma}^{gg}&=& \sigma_g^{(0)} \,\frac{\as}{2\pi}\,\left\{
   4 C_A \ln^2\bar{N}+
4C_A \frac{\ln\bar{N}}{N}+11+3\pi^2 \nonumber \right\}+{\cal O}
\left(\frac{\ln\bar{N}}{N^2} \right)\, , \nonumber
\\
\hat{\sigma}^{gq}&=&  \sigma_g^{(0)} \,\frac{\as}{2\pi}\,
C_F\; \frac{1}{N} \left[ -2
\ln\bar{N} +1\right]
+{\cal O}\left(\frac{\ln\bar{N}}{N^2} \right)
\; .
\end{eqnarray}
The full (``exact'') NLO cross total section has been calculated
in refs.~\cite{NLO}. When taking Mellin-$N$ moments of these
results, we obtain
\begin{eqnarray} \label{sec2}
\hat{\sigma}^{gg}_{\rm exact}&=& \sigma_g^{(0)} \,
\frac{\as}{2\pi}\,\left\{
4 C_A S_1^2(N) - 8C_A \left[ 
\frac{1}{N(N-1)}+\frac{1}{(N+1)(N+2)}\right]
   S_1(N)+11+4\pi^2\right. \nonumber \\
&&\left. +\frac{3 C_A}{N(N+1)}+\frac{13C_A}{(N-1)(N+2)}
   +4 C_A \left[\frac{1}{(N-1)^2}-\frac{1}{N^2}-\frac{1}{(N+1)^2}
+\frac{1}{(N+2)^2}\right]
\right\}
\quad , \nonumber \\
\hat{\sigma}^{gq}_{\rm exact}&=&  \sigma_g^{(0)} \,\frac{\as}{2\pi}\,
C_F\,\left\{ -2 S_1(N)\frac{N^2+N+2}{N(N+1)(N-1)}
   +\frac{2}{(N-1)^2} -\frac{2}{N^2} -\frac{1}{(N+1)^2}\right.
\nonumber \\
&&
+ \left.\;\frac{5}{2(N-1)}-\frac{1}{N}-\frac{1}{2(N+1)}\;\right\}\; .
\end{eqnarray}
At large $N$, this gives
\begin{eqnarray} \label{largen2}
\hat{\sigma}^{gg}&=& \sigma_g^{(0)} \,\frac{\as}{2\pi}\,\left\{
   4 C_A \ln^2\bar{N}+
4C_A \frac{\ln\bar{N}}{N}+11+4\pi^2 \right\}+
{\cal O}\left(\frac{\ln\bar{N}}{N^2} \right)\, , \nonumber
\\
\hat{\sigma}^{gq}&=&  \sigma_g^{(0)} \,\frac{\as}{2\pi}\,
C_F\; \frac{1}{N} \left[ -2
\ln\bar{N}+1\right]
+{\cal O}\left(\frac{\ln\bar{N}}{N^2} \right)
\; .
\end{eqnarray}
Comparing Eqs.~(\ref{largen1}) and (\ref{largen2}), we see that
they agree even to the subleading order ${\cal O}(1/N)$, except
for the contribution proportional to $\pi^2$ discussed in 
section 2.2. As noted there, had we fully developed the joint 
formalism to NNLL for {\em both} the threshold and the small-$Q_T$ 
limits, it would automatically account for this difference in the 
$C$ coefficients. At the level of our rudimentary inclusion of NNLL 
effects at large $b$, we need to accept the mismatch in the $\pi^2$ term. 

In addition, there are differences between the ${\cal O}(\as)$
expansion of the resummed cross section, Eq.~(\ref{sec1}),
and the exact ${\cal O}(\as)$, Eq.~(\ref{sec2}), that are
formally even more suppressed than the above $\pi^2$ term,
or even the ${\cal O}(\ln\bar{N}/N)$ terms mentioned earlier,
but that still have a fairly large impact on the results. These
correspond to the terms $\propto 1/(N-1)^k\;$ ($k=1,\, 2$) in these 
equations. For instance, there is a term $-8 C_A\ln\bar{N}/N(N-1)$
in Eq.~(\ref{sec1}) that arises from our treatment of evolution.
In Eq.~(\ref{sec2}), on the other hand, we find the combination
$-8 C_A S_1(N)/N(N-1)+13C_A/(N-1)(N+2)+4C_A/(N-1)^2$. 
Numerically, the former term leads to a sizeable negative 
contribution in Eq.~(\ref{sec1}), while the latter terms cancel 
each other to a large extent in Eq.~(\ref{sec2}).
This is the main reason why our jointly resummed cross section
integrates to a smaller area than one obtains for
the standard threshold-resummed cross section on the basis
of Eq.~(\ref{largen2}), where of course no $\propto 1/(N-1)^k$
contributions are present since they are suppressed from the
threshold point of view. 

The relevance of the $1/(N-1)^k$ contributions
results from the fact that these terms actually constitute
the most singular behavior in the limit opposite to 
partonic threshold, $z=Q^2/\hat{s}\to 0$. For Higgs masses 
of a few hundred GeV and LHC's center-of-mass
energy one is actually quite far from threshold. 
We note that at higher orders in $\as$ the leading terms singular 
at $N=1$ generalize to $\as^k/(N-1)^{2k}$, corresponding
to $\as^k\ln^{2k-1}(z)/z$ in $z$ space. These terms, too,
can be resummed to all orders in $\as$, and the resummation
was achieved in Ref.~\cite{Hautmann}. To what extent it would be 
possible to incorporate this small-$z$ resummation into our 
joint formalism is an interesting question that, however,
is beyond the scope of this paper. That the leading and 
subleading terms $\propto 1/(N-1)^k$  
($k=1,\, 2$) almost entirely cancel numerically
in the full NLO inclusive cross section, Eq.~(\ref{sec2}),
is rather remarkable, albeit probably not accidental. 
It is only because of this feature that the near-threshold 
approximation for the NLO cross section, Eq.~(\ref{largen2}), 
becomes a good approximation. 

Assuming that threshold resummation indeed does yield a 
reliable answer for the integrated total cross
section, it becomes then an interesting question what the implications
for the $Q_T$-dependence of the Higgs cross section will be, if it is
derived in such a way that it the area underneath the distribution
integrates to the threshold-resummed cross section. A possible, but 
clearly not unique, way of making up for the ``missing'' area is to 
redefine the $C$ coefficients of Eq.~(\ref{ccoeff}) in the following way:
\ba
\label{ccoeff1}
\tilde{C}_{g/g}\left( N,\as \right) &=& 1+
\frac{\as}{4\pi} \,\left\{ \,\pi^2+\frac{1}{1+\bar{b}^2}\left[
\frac{13C_A}{(N-1)(N+2)}
+\frac{4C_A}{(N-1)^2}\right] \right\}\; ,\nonumber \\
\tilde{C}_{g/q}\left( N,\as\right) &=&\frac{\as}{2\pi} C_F \,\left\{ \,
\frac{1}{N+1}+\frac{1}{1+\bar{b}^2}\,\frac{2}{(N-1)^2}\right\}\;  .
\ea
Adding terms of this form to the ${\cal C}$ coefficients is completely
consistent with resummation for both the threshold and
$Q_T$ resummation cases. The term $1/(1+\bar{b}^2)$ is designed so that
it becomes unity in the $Q_T$-integrated cross section and does 
not produce behavior singular in $Q_T$. We do not
claim that the coefficients in (\ref{ccoeff1}) are necessarily more 
realistic or better motivated than the standard ones of  Eq.~(\ref{ccoeff});
we rather regard them as one model for the effects associated 
with the $\propto 1/(N-1)^k$ terms on the Higgs $Q_T$ distribution, 
for which the $Q_T$-integrated cross section becomes numerically 
equivalent to a NLL threshold-resummed cross section. 

\begin{figure}[t!]
\begin{center}
\epsfig{file=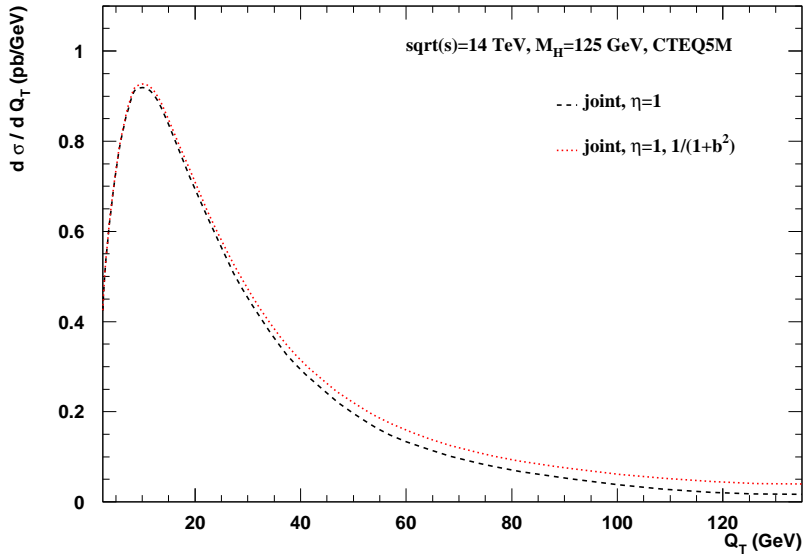,width=12cm}
\caption{Effect of introducing extra terms in the $C$ coefficients, as
shown in Eq.~(\ref{ccoeff1}) and discussed in the text, on the 
transverse momentum distribution for Higgs
production at the LHC in the framework of joint resummation.}
\label{higgs2}
\end{center}
\end{figure}

In Fig.~\ref{higgs2} we show the effect of the modified $C$ coefficients 
on the Higgs $Q_T$-distribution. As expected, there is no
difference between the $Q_T$ distributions with the $C$ coefficients in 
the forms~(\ref{ccoeff}) and~(\ref{ccoeff1}) in the region where
the distribution peaks; however, the new coefficients cause the 
$Q_T$ distribution to be higher at intermediate / large values of $Q_T$. 
The value of the total cross section is now about 41 pb.

\section{Conclusions}
In this paper we have presented a study of the transverse
momentum distribution of Higgs bosons in the context
of the joint resummation formalism. Guided by the
literature on the more standard $Q_T$ resummation,
we have further developed the joint resummation formalism 
to make it suitable for application to the Higgs production
cross section. This included the implementation of
some NNLL terms that have a significant numerical 
effect on the Higgs $Q_T$ distribution.

We have found in our study that effects associated
with threshold resummation have a modest
importance for the Higgs cross section at small
to moderate $Q_T$. For this region, where the 
distribution peaks, our results further confirm the applicability 
of standard $Q_T$ resummation.

On the other hand, threshold resummation is very
relevant for the $Q_T$-integrated cross section,
as is known from previous studies in the literature.
For joint resummation, the integrated cross
section is formally identical to the threshold
resummed one in the large-$N$ limit, but contains non-leading
contributions that go beyond standard threshold
resummation and which, as we found, decrease the cross
section. They are associated with {\em small}-$x$
terms from parton evolution, in a sense of
opposite origin to the threshold logarithms. 
We have argued that their presence indicates
that the Higgs $Q_T$ distribution at rather large
$Q_T$ receives sizeable contributions also from 
sources not related to Sudakov logarithms. Our study also suggests 
that it will be useful to extend joint resummation to NNLL.  
Such an extension will help clarify technical
issues, such as the choice of scales in coefficients and parton
distributions, which have a modest but significant effect on
the magnitude of the cross section even near the peak.

\section*{Acknowledgments}
We are grateful to D.\ de Florian and F.\ Hautmann for valuable discussions.
The work of G.S.\ was supported in part by the National Science 
Foundation, grants PHY9722101 and PHY0098527.
W.V.\ is grateful to RIKEN, Brookhaven National Laboratory and the U.S.
Department of Energy (contract number DE-AC02-98CH10886) for
providing the facilities essential for the completion of this work.
A.K.\ thanks the U.S. Department of Energy (contract number 
DE-AC02-98CH10886)
for financial support.

\begin{appendix}

\section{Leading- and next-to-leading logarithmic expansion}
Here we give the functions $h_g^{(0)}$ and $h_g^{(1)}$ that
determine the jointly resummed exponent~(\ref{expdef})
to leading and next-to-leading logarithmic accuracy~\cite{LSV,KSV}:
\begin{eqnarray}
h_g^{(0)} (\beta) &=& \frac{A_g^{(1)}}{2\pi b_0^2}
\left[ 2 \beta + \ln(1-2 \beta) \right]\, ,\\
h_g^{(1)} (\beta,Q,\mu) &=&
\frac{A_g^{(1)} b_1}{2\pi b_0^3} \left[ \frac{1}{2} \ln^2 (1-2 \beta) +
\frac{2 \beta + \ln(1-2 \beta)}{1-2\beta} \right] +
\frac{B_g^{(1)}}{2\pi b_0}  \ln(1-2 \beta) \nonumber \\
&+& \frac{1}{2\pi b_0} \left[ A_g^{(1)}\ln \left( \frac{Q^2}{\mu^2} 
\right)
-\frac{A_g^{(2)}}{\pi b_0}\right] \;
\left[ \frac{2 \beta}{1-2\beta}+ \ln(1-2 \beta) \right] \,,
\label{hsubadef}
\end{eqnarray}
with the coefficients $A_g^{(i)}$, $B_g^{(i)}$ from 
Eq.~(\ref{abcoeffs}),
$\beta$ from Eq.~(\ref{varsdef}), and $b_0$ as in Eq.~(\ref{b0def}).
In addition,
\ba \label{b1def}
b_1 \;=\; \frac{17 C_A^2-10 C_A T_R N_F-6 C_F T_R N_F}{24 \pi^2}\; .
\end{eqnarray}

\end{appendix}

\end{document}